\documentclass[prd,twocolumn,nofootinbib,preprintnumbers,amsmath,amssymb]{revtex4-2}

\usepackage{graphicx}
\usepackage{amsmath,amssymb}
\usepackage{bm}
\usepackage{color}
\usepackage[colorlinks=true,linkcolor=red,citecolor=blue,urlcolor=blue]{hyperref}
\usepackage{orcidlink}

\begin{document}

\title{
Comment on ``Quantum teleportation, entanglement, LQU and LQFI in $e^{+} e^{-} \rightarrow \mathrm{Y} \overline{\mathrm{Y}}$ processes at BESIII through noisy channels''
}

\author{Saeed~Haddadi\orcidlink{0000-0002-1596-0763}}
\email{haddadi@ipm.ir}
\affiliation{School of Particles and Accelerators, Institute for Research in Fundamental Sciences (IPM), P.O. Box 19395-5531, Tehran, Iran}

\date{\today}


\begin{abstract}
We provide a critical assessment of a recent study applying quantum information concepts, including noisy channels and teleportation fidelity, to hyperon–antihyperon pairs produced in $e^{+}e^{-} \to Y\bar Y$ reactions at BESIII. While the spin density matrix reconstructed from experimental data provides a physically meaningful description of production correlations, we argue that its subsequent interpretation in terms of standard decoherence models—such as amplitude damping, phase damping, and phase flip—lacks a clear physical correspondence for these systems. The produced particles emerge from a single scattering event and propagate as free, unstable relativistic states, without a well-defined system–environment interaction acting on their spin degrees of freedom. As a result, the variation of quantum correlations with an abstract noise parameter does not describe a genuine physical evolution. We further contend that the reported teleportation fidelity should not be interpreted as evidence for operational quantum communication, since hyperon states cannot be prepared, controlled, or measured in a way that would enable a realizable teleportation protocol. More generally, quantities such as logarithmic negativity, local quantum uncertainty, and local quantum Fisher information primarily characterize static production correlations rather than directly usable quantum resources. Our analysis highlights the importance of distinguishing between formal quantum-information measures and their physical interpretation in high-energy particle systems.
\end{abstract}

\maketitle


Recently, E.~Jaloum and M.~Amazioug~\cite{Ref1} investigated quantum teleportation fidelity and several measures of quantum correlations, including logarithmic negativity, local quantum uncertainty, and local quantum Fisher information, for hyperon–antihyperon pairs produced in $e^{+}e^{-}$ annihilation at BESIII. Their analysis further incorporates standard quantum noise models and discusses the robustness of quantum correlations under such channels.

Before proceeding, we emphasize that the spin density matrix reconstructed from experimental data provides a physically meaningful and well-established characterization of hyperon–antihyperon production. Our critique is not directed at this reconstruction, nor at the formal evaluation of quantum-information-theoretic quantities based on it. Rather, it concerns the physical interpretation of these quantities in terms of dynamical noisy channels and operational quantum-information protocols.

In standard quantum information theory, noise channels such as amplitude damping, phase damping, and phase flip are described by completely positive trace-preserving (CPTP) maps acting on well-defined subsystems, typically arising from controlled or well-characterized system–environment interactions~\cite{Ref3,Ref4}. In the case of hyperon–antihyperon pairs, however, the particles are produced in a localized scattering event and subsequently propagate as free, unstable relativistic states. There is no identified environment that couples specifically to their spin degrees of freedom in a manner that generates a CPTP map of the standard form. Effects such as detector inefficiencies or external perturbations primarily influence measurement statistics rather than inducing a dynamical evolution of the underlying quantum state. Consequently, the introduction of such decoherence channels should be understood as a phenomenological modeling assumption rather than a physically established dynamical process.

It is also important to note that the weak decay of hyperons corresponds to a non–trace-preserving process and therefore cannot be represented as a standard CPTP quantum channel acting solely on the spin degrees of freedom.

The interpretation of teleportation fidelity in this context requires careful consideration. In standard quantum information theory, teleportation protocols rely on the ability to prepare, manipulate, and measure quantum states under controlled conditions, including joint measurements and classical communication. In the hyperon system, these requirements are not satisfied: the particles are not prepared on demand, cannot be coherently controlled or stored, and do not permit adaptive measurement strategies. Therefore, while the calculated fidelity is mathematically well-defined for the reconstructed state, it should be regarded as a formal quantity rather than evidence of a physically realizable teleportation protocol.

Similarly, the angular dependence of logarithmic negativity, local quantum uncertainty, and local quantum Fisher information reflects the structure of spin correlations determined by the production mechanism and underlying conservation laws, rather than a dynamical process driven by environmental decoherence.

More generally, care must be taken when interpreting mathematical quantifiers of quantum correlations as operational quantum resources. While such measures are useful for characterizing properties of quantum states, attributing to them a direct operational role in communication or information processing requires physical controllability that is not available in the present system.

In conclusion, Ref.~\cite{Ref1} presents the application of quantum-information-theoretic tools to hyperon–antihyperon systems. However, the interpretation of the results in terms of standard noisy channels and operational quantum protocols requires careful qualification. In particular, the use of canonical decoherence models, the interpretation of teleportation fidelity, and the treatment of quantum correlations as directly usable resources are not supported by an established physical framework for these systems. We emphasize the importance of clearly distinguishing between formal quantum-information analyses of reconstructed density matrices and physically realizable quantum processes in high-energy experiments.


\end{document}